\documentclass[twocolumn,prb,amsmath,amssymb,floatfix]{revtex4}

\usepackage{graphics}
\usepackage{graphicx}
\usepackage{dcolumn}
\usepackage{bm}
\usepackage{subfigure}
\usepackage{slashed}
\usepackage{tensor}
\usepackage{bbold}
\usepackage{feynmf} 

\def\Tr{\mbox{Tr}}

\def\al{\alpha}

\def\be{\begin{equation}}
\def\ee{\end{equation}}
\def\bea{\begin{eqnarray}}
\def\eea{\end{eqnarray}}
\def\bse{\begin{subequations}}
\def\ese{\end{subequations}}
\def\bc{\begin{center}}
\def\ec{\end{center}}

\def\nonum{\nonumber}

\def\E{{e}}
\def\I{{i}}

\def\D{{d}}
\def\Amu{{A_\mu}}
\def\Anu{{A_\nu}}
\def\Flmn{{F_{\mu \nu}}}

\def\Bpsi{{{\bar{\psi}}}}

\def\Sp{{\slashed p}}
\def\Sq{{\slashed q}}
\def\Sk{{\slashed k}}
\def\Dmu{{\partial_\mu}}
\def\Dnu{{\partial_\nu}}
\def\DDmu{{D_\mu}}

\def\Lag{{\mathcal{L}}}

\newcommand*\pFqskip{8mu}
\catcode`,\active
\newcommand*\pFq{\begingroup
        \catcode`\,\active
        \def ,{\mskip\pFqskip\relax}%
        \dopFq
}
\catcode`\,12
\def\dopFq#1#2#3#4#5{%
        {}_{#1}F_{#2}\biggl(\genfrac..{0pt}{}{#3}{#4} \bigg | #5\biggr)%
        \endgroup
}

\begin{document}

\title{Electromagnetic current correlations in reduced quantum electrodynamics}

\author{S. Teber}
\email{teber@lpthe.jussieu.fr}
\affiliation{Laboratoire de Physique Th\'eorique et Hautes Energies, Universit\'e Pierre et Marie Curie, 4 place Jussieu, 75005, Paris, France}

\date{\today}

\begin{abstract}
We consider a theory of massless reduced quantum electrodynamics (RQED$_{d_\gamma,d_e}$), e.g., a quantum field theory where the $U(1)$ gauge field
lives in $d_\gamma$-spacetime dimensions while the fermionic field lives in a reduced spacetime of $d_e$ dimensions ($d_e \leqslant d_\gamma$). 
In the case where $d_\gamma=4$ such RQEDs are renormalizable while they are super-renormalizable for $d_\gamma <4$. 
The 2-loop electromagnetic current correlation function is computed exactly for a general RQED$_{d_\gamma,d_e}$. 
Focusing on RQED$_{4,3}$, the corresponding $\beta$-function is shown to vanish which implies the scale invariance of the theory. 
Interaction correction to the 1-loop vacuum polarization, $\Pi_1$, of RQED$_{4,3}$ is found to be:
$\Pi = \Pi_1 \,(1 + 0.056 \,\al)$ where $\al$ is the fine structure constant.
The scaling dimension of the fermion field is computed at 1-loop and is shown to be anomalous for RQED$_{4,3}$.
\end{abstract}

\maketitle

\section{Introduction}
\label{Sec.Introduction}

Reduced quantum electrodynamics (RQED$_{d_\gamma,d_e}$) [\onlinecite{Gorbar2001}] is a quantum field theory (QFT) describing the interaction of an abelian $U(1)$ gauge field
living in $d_\gamma$ space-time dimensions with a fermion field living in a reduced space-time of $d_e$ dimensions ($d_e \leqslant d_\gamma$).
In the particular case where gauge and fermion fields live in the same space-time, $d_\gamma=d_e$, RQEDs correspond to the usual QEDs. 
QED$_4$ is the simplest and most well-established renormalizable QFT which presently exists, see
textbooks~[\onlinecite{QFTTextBooks}]. In dimensions lower than $3+1$, such theories are super-renormalizable. 
QED$_2$, also known as the Schwinger model, is a celebrated exactly solvable model~[\onlinecite{Schwinger62}]. QED$_3$ was considered as
a toy model to study confinement~[\onlinecite{Feynman81}], infrared divergences~[\onlinecite{QED3_IR_papers}] and chiral symmetry breaking~[\onlinecite{Appelquist86}]. 
In condensed matter physics, non-relativistic $2+1$-dimensional gauge field theories away from the charge neutral point, where the gauge field is generated by site occupation constraint,
are used to model copper oxides, see e.g., Refs.~[\onlinecite{MarstonAffleck89,IoffeWiegmann90}] and references therein.

Motivations for the study of reduced theories came from interest in branes~\cite{Gorbar2001,Dimopoulos2001}
as well as potential applications to condensed matter physics, see Refs.~[\onlinecite{Kovner90+Dorey92}] for early studies related to planar superconductivity. 
It is indeed quite usual that non-relativistic electrons may be restricted to a plane ($d_e=2+1$), e.g., 
graphene, or a line ($d_e=1+1$), e.g., quantum wires, while interacting through a bulk gauge field ($d_\gamma = 3+1$). 
In the case of graphene, see [\onlinecite{GrapheneReview,Semenoff2011}] for reviews, low-energy excitations are massless Dirac fermions with a Fermi velocity much smaller than the velocity of 
light, $v_F/c \approx 1/300$, and interacting via the long-range instantaneous Coulomb interaction. The strength of the later is measured by a 
fine structure constant, $\al_g = e^2 / 4 \pi \epsilon \hbar v_F$, which may be of the order of unity unless the substrate is a high-$\kappa$ dielectric.
Nevertheless, electronic properties of graphene were computed via perturbation theory and, in some cases, different results were found
[\onlinecite{HerbutMishchenko2008}]. Moreover, it has been claimed that the system flows towards a Lorentz covariant fixed point in the 
infrared, cf., [\onlinecite{Vozmediano2011}] for a recent review.

In the present paper, we compute radiative corrections in massless RQED$_{d_\gamma,d_e}$ with a special focus on
electromagnetic current correlations. The derived formulas are applied to the case of RQED$_{4,3}$ which is an ultra-relativistic version of graphene, i.e., pair creation
is restricted to a plane. Feynman diagrams with non-integer indices appear which originate from the fact that the reduced theory is non-local. 
The diagrams are computed using advanced multi-loop techniques~[\onlinecite{GrozinBook,SmirnovBook,Kotikov96}]. 
In the following we will follow the notations of Ref.~[\onlinecite{GrozinBook}].

The paper is organized as follows. In Sec.~\ref{Sec.Model}, the model of massless RQED$_{d_\gamma,d_e}$ is presented. 
In Sec.~\ref{Sec.Rules}, the reduced Feynman rules and the renormalization scheme are presented.
In Sec.~\ref{Sec.1loop}, 1-loop corrections to vacuum polarization and fermion self-energy are presented and applied to RQED$_{4,3}$.
In Sec.~\ref{Sec.2loop}, 2-loop corrections to the vacuum polarization are presented and applied to RQED$_{4,3}$. 
Finally, in Sec.~\ref{Sec.Conclusion} we summarize our results and conclude.
In the following, we work in units where $\hbar=c=1$.

\section{Model}
\label{Sec.Model}

Massless RQED$_{d_\gamma,d_e}$ is described by:
\bea
S_{d_\gamma,d_e} =&& \int \D^{d_\gamma} x \, \Lag_{d_\gamma,d_e},
\label{rqed} \\
\Lag_{d_\gamma,d_e}\,=&&\,\Bpsi(x) \I \gamma^{\mu_e}D_{\mu_e} \psi(x)\,\delta^{(d_\gamma-d_e)}(x) -
 \nonum \\
&&-\frac{1}{4}\,F_{\mu_\gamma \nu_\gamma}F^{\mu_\gamma \nu_\gamma} - \frac{1}{2a}\left(\partial_{\mu_\gamma}A^{\mu_\gamma}\right)^2,
\nonum
\eea
where $\Flmn = \Dmu \Anu - \Dnu \Amu$ is the field stress tensor, $\DDmu = \Dmu + \I \E \Amu$ is the gauge covariant derivative and $a$ is a gauge fixing parameter.
Eq.~(\ref{rqed}) displays two kinds of indices where the zero index refers to time. 
Matter indices: $\mu_e\,=\,0,\,1,\,...,\,d_e-1$, are related to the first term in the Lagrangian which is a boundary term describing a fermion field $\psi$.
Gauge indices: $\mu_\gamma \,=\, 0,\,1,\,...,\,d_e-1,\,d_e,\,...,\,d_\gamma-1$, are related to the bulk gauge field $\Amu$. 
The minimal coupling of the gauge field to the fermion current, $j_\mu A^\mu$, only involves fermionic indices so that
the conserved current can be defined as:
\be
j^\mu(x) =
\left\{
          \begin{array}{ll}
                \E\, \Bpsi(x) \gamma^\mu \psi(x) \delta^{(d_\gamma-d_e)}(x) & \,\,\, \mu=\mu_e, \\
                0 & \,\,\, \mu = d_e,\,...,\,d_\gamma-1. \\
          \end{array}
\right.
\label{fc} 
\ee

For general $d_e$ and $d_\gamma$ the dimensions of the fields are given by:
\begin{subequations}
\label{dims+params}
\bea
[A_\mu] &=& 1 - \varepsilon_\gamma,
\quad [\psi]= \frac{3}{2} - \varepsilon_e - \varepsilon_\gamma,
\quad [e] = \varepsilon_\gamma,
\label{dims}
\\ 
&&\varepsilon_\gamma = \frac{4-d_\gamma}{2}, 
\qquad \varepsilon_e = \frac{d_\gamma - d_e}{2},
\label{params}
\eea
\end{subequations}
where the variables $\varepsilon_\gamma$ and $\varepsilon_e$ will be useful later on.
From Eq.~(\ref{dims+params}), we see that for a $4$-dimensional gauge field ($d_\gamma=4$) the coupling constant is dimensionless whatever space
the fermion field lives in. This includes the standard case of QED$_4$ ($d_e=d_\gamma=4$) but also RQED$_{4,3}$ ($d_\gamma=4$ and $d_e=3$) 
and RQED$_{4,2}$ ($d_\gamma=4$ and $d_e=2$) and suggests that all these theories are renormalizable quantum field theories. 
This is in agreement with the counting of ultraviolet (UV) divergences.
Indeed, the superficial degree of divergence (SDD) of an RQED$_{d_\gamma,d_e}$ diagram is given by:
\be
D = d_e + \frac{d_\gamma - 4}{2} V - \frac{d_\gamma-2}{2} N_\gamma -\frac{d_e-1}{2}N_e,
\label{sdd}
\ee
where $V$ is the number of vertices, $N_\gamma$ the number of external gauge lines and $N_e$ the number of external
fermion lines. From Eq.~(\ref{sdd}) we see that for $d_\gamma=4$ the SDD does not depend on the number of vertices whatever
value $d_e$ takes. This is in marked contrast with super-renormalizable theories which have coupling constants with dimensions of positive power of the mass. 
Going to higher orders, one accumulates negative powers of momentum in order to compensate for the dimensionality of the coupling constant.
Such theories are therefore UV finite. Infrared (IR) divergences may however force the breakdown of the loop expansion for massless theories, 
cf., Refs.~[\onlinecite{QED3_IR_papers}] for the case of QED$_3$.

Concerning RQEDs, we see from Eq.~(\ref{sdd}) that amongst the most superficially divergent amplitudes, the fermion self-energy ($N_e=2$, $N_\gamma=0$) 
and the fermion-gauge vertex ($N_e=2$, $N_\gamma=1$) of all RQED$_{4,d_e}$s have the same SDD as in QED$_4$: $D_e = 1$ and $D_{e-\gamma}=0$, respectively. 
Because the fermion self-energy may only depend on $\slashed p$, the degree of divergence is actually reduced to $D_e=0$ so that both of 
these amplitudes are logarithmically divergent in RQEDs. 

On the other hand, the SDD of the photon self-energy ($N_e=0$, $N_\gamma=2$) is reduced in RQEDs: $D_\gamma=1$ for RQED$_{4,3}$ and $D_\gamma=0$ for RQED$_{4,2}$, with respect 
to QED$_4$ (where $D_\gamma=2$). Because of current conservation, this electromagnetic current correlation function: $\Pi^{\mu \nu}(x-y) = \langle j^\mu(x) j^\nu(y) \rangle$
where the current is given by Eq.~(\ref{fc}), is constrained to have the general form:
\be
\Pi^{\mu \nu}(q) = \big(\, g^{\mu \nu} q^2 - q^\mu q^\nu\,\big)\,\Pi(q^2),
\label{pimunu}
\ee
where $g^{\mu \nu} = {\rm diag}(1,-1,-1,-1)$ is the metric tensor and $q^2 =q_\mu q^\mu$.
This constraint further reduces the degree of divergence to: $D_\gamma=0$ for QED$_4$, $D_\gamma=-1$ for RQED$_{4,3}$ 
and $D_\gamma=-2$ for RQED$_{4,2}$. This dimensional analysis therefore suggests that while $\Pi(q^2)$ 
logarithmically diverges in QED$_4$ it is finite in RQEDs.

The fact that the coupling constant of massless RQED$_{4,d_e}$ is dimensionless 
implies scale invariance at the classical level, see Ref.~[\onlinecite{ColemanBook}] for a review.
Scale invariance may however be broken by quantum corrections~\cite{ColemanBook} if the latter diverge order by order in perturbation theory. 
This is because a cut-off has to be introduced in order to remove the divergences. 
Scale invariance is for example lost in QED$_4$ where the coupling constant diverges in the ultraviolet. 
This is encoded in the beta function of the theory which describes the dependence of the coupling constant on the energy scale:
\be
\beta(\al(\mu)) = \frac{d \log \al(\mu)}{d \log \mu},
\label{beta_def}
\ee
and $\beta > 0 $ for QED$_4$.
The above dimensional analysis suggests that massless RQED$_{4,d_e}$, with $d_e<4$, may remain scale invariant when radiative corrections are included with a possible anomalous
scaling dimension for the fermion field. We will check this in the following by an explicit computation of the 1-loop 
anomalous scaling dimension of the fermion field and of the 2-loop beta function. Results will be applied to RQED$_{4,3}$.

\section{Feynman rules and renormalization scheme}
\label{Sec.Rules}

\begin{figure}
    \includegraphics{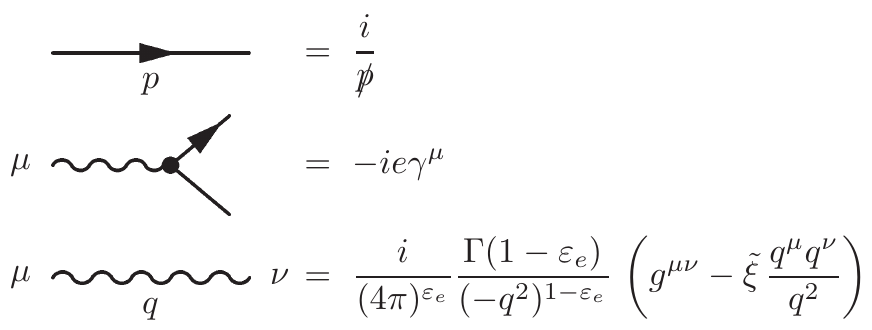}
    \caption{\label{fig:FeynmanRules}
    Feynman rules for massless RQED.}
\end{figure}

Keeping the space-time dimensions of the fermion field, $d_e$, and gauge-field, $d_\gamma$, arbitrary for the moment, we proceed on deriving the Feynman rules of
massless RQED$_{d_\gamma,d_e}$. The free massless fermionic propagator reads:
\be
S_0(p_e) = \frac{i\Sp_e}{p_e^2 + i\epsilon},
\label{mprop0}
\ee
where $p_e=p_0,\,...,\,p_{d_e-1}$ lies in the reduced matter space and $i\epsilon=i0^+$ is the convergence factor of this Feynman-type propagator. 
On the other hand, the free bulk gauge field propagator reads:
\be
D_0^{\mu_\gamma \nu_\gamma}(q_\gamma) = \frac{-i}{q_\gamma^2+i\varepsilon}\left( g^{\mu_\gamma \nu_\gamma} - \xi \, \frac{q_\gamma^{\mu_\gamma} q_\gamma^{\nu_\gamma}}{q_\gamma^2+i\varepsilon} \right),
\label{gprop0}
\ee
where $\xi = 1-a$ and $q_\gamma=q_0,\,...,\,q_{d_\gamma-1}$ lives in the bulk $d_\gamma$-dimensional space-time. 
In the following we will often omit the convergence factor in order to simplify notations but they are implied. 
Because we are interested in the properties of the system in the reduced space where the fermion field is living we
may integrate over the $d_\gamma - d_e$ bulk gauge degrees of freedom. This yields an effective
reduced gauge field propagator:
\be
\tilde{D}_0^{\mu_e \nu_e}(q_e) = \frac{i}{(4\pi)^{\varepsilon_e}}\frac{\Gamma(1-\varepsilon_e)}{(-q_e^2)^{1-\varepsilon_e}}\,\left( g^{\mu_e \nu_e} - \tilde{\xi}\,\frac{q_e^{\mu_e} q_e^{\nu_e}}{q_e^2}\right),
\label{rgprop0}
\ee
where $\tilde{\xi} = \xi (1-\varepsilon_e)$ and $\varepsilon_e$ was defined in Eq.~(\ref{params}).
This propagator can be separated in longitudinal and transverse parts which read:
\bea
&&\tilde{D}_{0 \parallel}(q^2) = \frac{i}{(4\pi)^{\varepsilon_e}}\frac{\Gamma(1-\varepsilon_e)}{(-q^2)^{1-\varepsilon_e}}\, \tilde{a},
\nonum \\
&&\tilde{D}_{0 \bot}(q^2) = \frac{i}{(4\pi)^{\varepsilon_e}}\frac{\Gamma(1-\varepsilon_e)}{(-q^2)^{1-\varepsilon_e}}.
\label{rgprop1}
\eea
In the case of RQED$_{4,3}$, $\varepsilon_e=1/2$, the reduced propagator has a square root branch-cut whereas for 
RQED$_{4,2}$, $\varepsilon_e=1$, it is logarithmically divergent. The latter may be cured by giving a small width to the wire~[\onlinecite{Gorbar2001}].
In both cases the reduced QFT is non-local. In the following we will focus on general formulas directly applicable to the cases of QED$_4$, QED$_3$ and RQED$_{4,3}$.
Feynman rules for these massless QEDs are summarized in Fig.~\ref{fig:FeynmanRules}.

Perturbation theory is implemented in the usual way by considering the
Dyson equation for the fermion and gauge field propagators. The former reads:
\be
S(p) = S_0(p) + S_0(p) \left( -i \Sigma(p) \right) S(p),
\ee
where it is understood that $p$ is in the fermion space and $\Sigma$ is the fermion self-energy.
The solution of this equation reads:
\be
S(p) = \frac{\I}{\Sp}\frac{1}{1-\Sigma_V(p^2)},
\label{mprop}
\ee
where the expression of the fermion self-energy suited to the massless case has been adopted~\cite{GrozinBook}:
\be
\Sigma(p) = \Sp \Sigma_V(p^2).
\label{sigma}
\ee
On the other hand, the Dyson equation for the reduced gauge field propagator reads:
\bea
\tilde{D}^{\mu_e \nu_e}(q_e) =
\tilde{D}_0^{\mu_e \nu_e}(q_e) +
\tilde{D}_0^{\mu_e \sigma_e}(q_e)\,i\Pi_{\sigma_e \rho_e}(q_e)\,\tilde{D}^{\rho_e \nu_e}(q_e).
\nonum
\eea
Iterating this equation, using Eqs.~(\ref{pimunu}) and (\ref{rgprop0}) and dropping sub-indices to simplify notations yields:
\bea
\tilde{D}^{\mu \nu}(q) = \tilde{D}_{\parallel}(q) \frac{q^{\mu} q^{\nu}}{q^2} +\tilde{D}_{\bot}(q)\left(g^{\mu \nu} - \frac{q^{\mu} q^{\nu}}{q^2}\right),
\nonum
\eea
where only the transverse part gets corrections:
\bea
&&\tilde{D}_{\parallel}(q^2) = \tilde{D}_{0 \parallel}(q^2),
\nonum \\ 
&&\tilde{D}_{\bot}(q^2) = \tilde{D}_{0 \bot}(q^2) \frac{1}{1- \tilde{D}_{0\bot}(q^2)\I q^2 \Pi(q^2)}.
\label{rgprop-d}
\eea
Finally, the dressed vertex reads:
\be
-ie\Gamma^\mu (p,p') = -ie\gamma^\mu -ie \Lambda^\mu (p,p'),
\ee
where $\Lambda^\mu (p,p')$ includes all radiative corrections.

In order to make sense of eventual diverging integrals in the perturbation series we will use dimensional regularization~[\onlinecite{thooft72}]. 
Loop integrals will depend on $d_e$ which may be expressed as a function of $\varepsilon_\gamma$ and $\varepsilon_e$, 
$d_e = 4 - 2\varepsilon_\gamma - 2\varepsilon_e$ from Eq.~(\ref{dims+params}). 
After computing the loop integrals for arbitrary values of the parameters we will fix $\varepsilon_e$, e.g.,
$\varepsilon_e=1/2$ for RQED$_{4,3}$, and perform an expansion in $\varepsilon_\gamma$, 
e.g., $\varepsilon_\gamma \rightarrow 0$ for RQED$_{4,3}$. 
Diverging Feynman diagrams then appear as meromorphic functions in $\varepsilon_\gamma$.
Subtracting divergences corresponds to subtracting poles in $1/\varepsilon_\gamma$. This can be done by relating the bare fields 
and parameters of the theory to renormalized quantities as: 
\be
\psi = Z_{\psi}^{1/2}\psi_r, \quad \tilde{A} = Z_A^{1/2} \tilde{A}_r, \quad \tilde{a}=Z_A\tilde{a}_r, \quad e=Z_\al^{1/2}e_r,
\label{ren_constants}
\ee
where $\tilde{A}$ is a reduced gauge field and $Z_\psi$, $Z_A$ and $Z_\al$ are dimensionless renormalization constants. 
We shall be using an improved minimal subtraction scheme ($\overline{\rm{MS}}$) where these constants reduce to unity in a finite QFT 
and take the form of Laurent series in $\varepsilon_\gamma$ in the presence of divergences.
Moreover, in order to restore the canonical dimensions of the fields and parameters one needs to introduce a renormalization scale
$\mu$ which has dimension of mass. In the $\overline{\rm{MS}}$ scheme we therefore define a dimensionless renormalized coupling constant $\al$ via the equation:
\be
\frac{\al(\mu)}{4\pi} = \mu^{-2\varepsilon_\gamma}\,\frac{e^2}{(4\pi)^{d_\gamma/2}}\,Z_\al^{-1}(\al(\mu))\,e^{-\gamma \varepsilon_\gamma},
\label{ren_coupling}
\ee
where the $\mu^{-2\varepsilon_\gamma}$ factor compensates for the dimension of $e^2$, Eq.~(\ref{dims+params}). 
Equivalently, from Eq.~(\ref{ren_coupling}), the bare coupling constant $e$ can be expressed via the renormalized coupling constant 
$\al(\mu)$.

From the above arguments, the relation between the bare and renormalized fermion propagators reads:
\be
S(p) = Z_\psi(\al(\mu),a_r(\mu)) S_r(p\,;\mu),
\label{mpropr}
\ee
where all singularities are in $Z_\psi$ and $S_r$ is finite.
Similarly, the gauge field propagator:
\be
\tilde{D}_{\mu \nu}(q) = Z_A(\al(\mu)) \tilde{D}_{\mu \nu}^r(q\,;\mu),
\ee
separates into a renormalized longitudinal part which is all contained in the renormalization of the
effective gauge fixing term: $\tilde{a}_r(\mu) = Z_A^{-1}(\al(\mu))\,\tilde{a}$,
and a renormalized transverse part which may receive radiative corrections:
$\tilde{D}_{\bot}^r(q^2\,;\mu) = Z_A^{-1}(\al(\mu))\,\tilde{D}_{\bot}(q^2)$.
 
Finally, one may introduce another constant, $Z_\Gamma$, for the vertex: $\Gamma^\mu = Z_\Gamma \Gamma^\mu_r$
where $\Gamma^\mu_r$ is finite.
Because both renormalized electron charge and vertex are finite there is a constraint among
the constants:
\be
Z_\al = (Z_\Gamma Z_\psi)^{-2} Z_A^{-1}.
\label{Zs-constraint}
\ee
In the following we will compute these constants and deduce the anomalous scaling dimensions of the fields:
\be
\gamma_i(\al(\mu),a_r(\mu)) = \frac{d \log Z_i(\al(\mu),a_r(\mu))}{d\log \mu}.
\label{asd}
\ee

\section{Perturbation theory: one-loop}
\label{Sec.1loop}

The results we shall derive in this section are expressed in terms of a one-loop massless propagator diagram~[\onlinecite{GrozinBook}]:
\bea
&&\int \frac{\D^{d_e}\,k}{(2\pi)^{d_e}}\,\frac{1}{D_1^{\nu_1} D_2^{\nu_2}} = \frac{i}{(4\pi)^{d_e/2}} (-q^2)^{\frac{d_e}{2}-\nu_1-\nu_2} G(\nu_1,\nu_2),
\nonum \\ 
&&D_1 = -(k+q)^2, \qquad D_2 = -k^2,
\label{1loopmassless}
\eea
where $\nu_1$ and $\nu_2$ are arbitrary indices labeling the diagram, the power of $-q^2$ follows from dimensionality 
and $G(\nu_1,\nu_2)$ is a dimensionless function which reads:
\be
G(\nu_1,\nu_2) = \frac{\Gamma(-d_e/2+\nu_1+\nu_2)\,\Gamma(d_e/2-\nu_1)\,\Gamma(d_e/2-\nu_2)}{\Gamma(\nu_1)\,\Gamma(\nu_2)\,\Gamma(d_e-\nu_1-\nu_2)},
\label{1loopG}
\ee
or, equivalently, in terms of $\varepsilon_\gamma$ and $\varepsilon_e$ by using $d_e = 4 - 2\varepsilon_\gamma - 2\varepsilon_e$ from Eq.~(\ref{dims+params}).
Straightforward dimensional analysis shows that a pole in the first gamma function of the numerator of $G(\nu_1,\nu_2)$ signals an ultraviolet divergence while a pole in
the two other gamma functions of the numerator signals an infrared divergence.

\begin{figure}
  \includegraphics{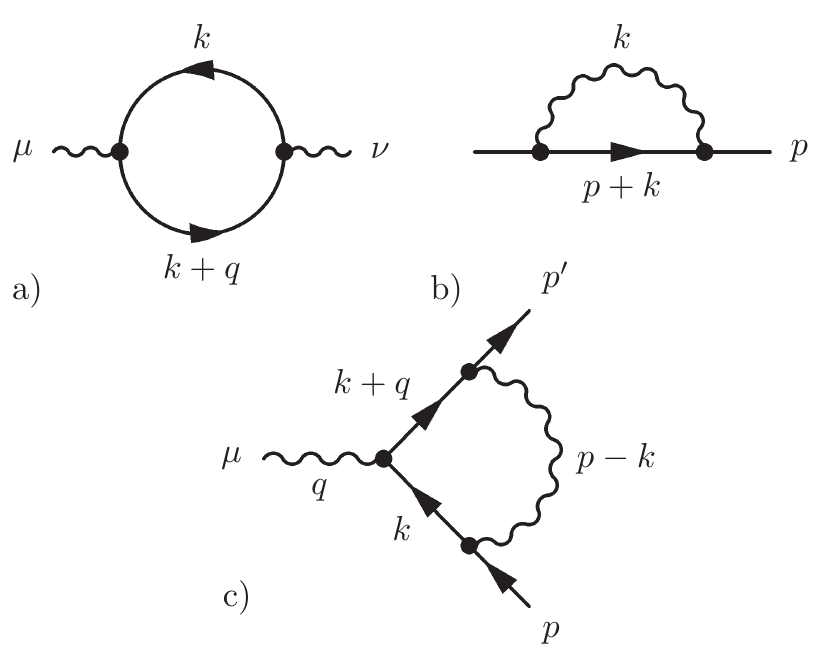}
  \caption{\label{fig:one-loop}
  One-loop diagrams: a) gauge field self-energy, b) fermion self-energy and c) fermion-gauge field vertex.}
\end{figure}

The one-loop diagrams, see Fig.~\ref{fig:one-loop}, are defined by:
\begin{widetext}
\begin{subequations}
\label{1loopdiagrams}
\bea
&&i\Pi_1^{\mu \nu}(q) = - \int \frac{\D^{d_e}k}{(2\pi)^{d_e}}\, \Tr \left[ (-ie\gamma^\mu) \frac{i(\Sk +\Sq)}{(k+q)^2}(-ie\gamma^\nu)\frac{i\Sk}{k^2} \right],
\label{pimunu-1l}
\\
&&\Sigma_1(p) = \int \frac{\D^{d_e}k}{(2\pi)^{d_e}}\, (-ie\gamma^\mu) \frac{i(\Sp +\Sk)}{(p+k)^2}(-ie\gamma^\nu)
\frac{i}{(4\pi)^{\varepsilon_e}}\frac{\Gamma(1-\varepsilon_e)}{(-k^2)^{1-\varepsilon_e}}
\left( g_{\mu \nu} - \tilde{\xi}\,\frac{k_\mu k_\nu}{k^2} \right),
\label{sigma-1l}
\\
&&-ie\Lambda_1^\mu = \int \frac{\D^{d_e}k}{(2\pi)^{d_e}}\, \frac{i}{(4\pi)^{\varepsilon_e}}\frac{\Gamma(1-\varepsilon_e)}{(-k^2)^{1-\varepsilon_e}}
\left( g^{\nu \rho} - \tilde{\xi}\,\frac{k^\nu k^\rho}{k^2} \right)(-ie \gamma^\nu) \frac{i \Sk}{k^2} (-ie\gamma^\mu) \frac{i \Sk}{k^2} (-ie \gamma^\rho),
\label{vertex-1l}
\eea
\end{subequations}
\end{widetext}
where, for our purposes, it is enough to compute the vertex correction for $p=p'=0$.
In order to perform the trace one has to specify the dimensionality of gamma matrices. For even $d_e$ the dimensionality of the gamma matrices is fixed and equals $d_e$.
On the other hand, for odd $d_e$ there is an ambiguity. In particular, for $d_e=3$ we may either consider a $2$-dimensional or a $4$-dimensional representation of the
gamma matrices, see e.g., Ref.~[\onlinecite{Appelquist86}]. For the sake of generality, we will work with $d$-dimensional gamma matrices.
This will also give to us a control over the final calculations. Indeed, the parameter $d$ is equivalent to a number of fermion species.
Going to large $d$ is equivalent to performing a large $N$ approximation in which correlation functions significantly simplify, cf., Ref.~[\onlinecite{LargeN}] for a review. 
With these conventions, some useful trace identities read:
\begin{widetext}
\begin{subequations}
\label{traceids}
\bea
&&\tensor{g}{^\mu_\mu}=\tensor{\delta}{^\mu_\mu}=d_e,
\qquad \Tr \left[ \mathbb{1} \right] = d,
\qquad
\Tr \left[ \gamma^\mu \gamma^\nu \right] = d\tensor{g}{^\mu^\nu},
\qquad\Tr \left[ \gamma^\mu \gamma^\al \gamma_\mu \gamma^\beta \right] = -d(d_e-2)\tensor{g}{^\al^\beta},
\label{trace_id}\\
&&\Tr \left[ \gamma_\mu \gamma^\al \gamma_\nu \gamma^\beta \gamma^\mu \gamma^\delta \gamma^\nu \gamma^\theta \right] = dg^{\al \beta} g^{\delta \theta} (4-d_e)(d_e-2)
-d g^{\al \delta} g^{\beta \theta} (8 - d_e)(d_e-2) +d g^{\al \theta}g^{\delta \beta}(4-d_e)(d_e-2).
\label{trace_id-2loop}
\eea
\end{subequations}
\end{widetext}
Such identities can be applied to Eq.~(\ref{pimunu}) in order to relate
$\Pi(q^2)$ to a contracted form of the gauge field self-energy:
\be
\Pi(q^2) = \frac{- \tensor{\Pi}{^\mu_\mu}(q)}{(d_e-1)(-q^2)}.
\label{pi_id}
\ee
Similarly, using Eq.~(\ref{sigma}):
\be
\Sigma_{V}(p^2) = \frac{1}{d}\,\Tr \left[ \Sp \Sigma(p) \right].
\ee

Using the above identities and computing the one-loop integrals yields, after some algebra:
\begin{widetext}
\begin{subequations}
\label{1loopdiagrams-1}
\bea
&&
\Pi_1(q^2) = - d\,\frac{e^2 \,\left(-q^2\right)^{-\varepsilon_\gamma-\varepsilon_e}}{(4\pi)^{d_e/2}}\frac{1-\varepsilon_\gamma -\varepsilon_e}{3- 2\varepsilon_\gamma -2\varepsilon_e}\,G(1,1),
\label{pimunu-1l-1}
\\
&&
\Sigma_{V1}(p^2) = -\frac{e^2 \,\Gamma(1-\varepsilon_e) (-p^2)^{-\varepsilon_\gamma}}{(4 \pi)^{d_\gamma/2}}
\left[ \frac{2(1-\varepsilon_\gamma-\varepsilon_e)^2}{2-2\varepsilon_\gamma-\varepsilon_e} - \xi\,(1-\varepsilon_\gamma-\varepsilon_e) \right]\,
G(1,1-\varepsilon_e), 
\label{sigma-1l-1}
\\
&&
\Lambda_1^\mu = \frac{e^2 \,\Gamma(1-\varepsilon_e) \, m^{-2\varepsilon_\gamma}}{(4\pi)^{d_\gamma/2}}\,\gamma^\mu\,
\left[ \frac{2(1-\varepsilon_\gamma -\varepsilon_e)^2}{2-\varepsilon_\gamma -\varepsilon_e} - \tilde{\xi} \,\right]\,
\frac{\Gamma(\varepsilon_\gamma)}{\Gamma(2\varepsilon_e)},
\label{vertex-1l-1}
\eea
\end{subequations}
\end{widetext}
where $G$ is the massless one-loop propagator of Eq.~(\ref{1loopG}), 
and a mass, $m$, has been introduced as an infrared regulator in the expression of the vertex correction.
Eqs.~(\ref{1loopdiagrams-1}) can be expressed in terms of the renormalized coupling constant by using 
Eq.~(\ref{ren_coupling}) at one-loop ($Z_\al = 1 + \mathcal{O}(\al)$). 

Eqs.~(\ref{1loopdiagrams-1}) apply to the well known cases of QED$_4$,
see Ref.~[\onlinecite{GrozinBook}], and QED$_3$, see Refs.~[\onlinecite{QED3_IR_papers}]. 
In order to facilitate the comparison with RQED$_{4,3}$ we give the 1-loop results of these well-known
theories. In the case of QED$_4$ we have to fix $\varepsilon_e=0$ and take the limit $\varepsilon_\gamma \rightarrow 0$ in 
Eqs.~(\ref{1loopdiagrams-1}). This yields:
\begin{subequations}
\label{qed4-gauge-1loop}
\bea
&&Z_A(\al) = 1 - \frac{d}{3}\,\frac{\al}{4\pi \varepsilon_\gamma} + \mathcal{O}\left(\al^2\right),
\label{qed4-ZA1} \\
&&D_{\bot}^r(q^2;\mu) = \frac{i}{-q^2} \left[1 + \frac{d}{3}\,\frac{\al}{4\pi}\,\left(L_q-\frac{5}{3}\right) + \mathcal{O}\left(\al^2\right) \right].
\label{qed4-D1} \\
&&\gamma_A(\al) = \frac{2d}{3}\,\frac{\al}{4\pi} + \mathcal{O}\left(\al^2\right),
\label{qed4-gA1}
\eea
\end{subequations}
where $L_q= \log \frac{-q^2}{\mu^2}$. Eq.~(\ref{qed4-gA1}) shows that the gauge field has an anomalous scaling dimension. Similarly:
\begin{subequations}
\label{qed4-fermion-1loop}
\bea
&& Z_\psi(\al,a) = 1 - \frac{\al a}{4\pi \varepsilon_\gamma} + \mathcal{O}\left(\al^2\right),
\label{qed4-Zpsi1} \\
&& Z_\Lambda(\al,a) = 1 + \frac{\al a}{4\pi \varepsilon_\gamma}\, + \mathcal{O}\left(\al^2\right),
\label{qed4-Zlambda1} \\
&& -i\Sp S_r(p;\mu) = 1 + \frac{\al a}{4\pi}\,\left(L_p-1\right) + \mathcal{O}\left(\al^2\right),
\label{qed4-S1} \\
&& \gamma_\psi(\al,a) = 2a\,\frac{\al}{4\pi} + \mathcal{O}\left(\al^2\right),
\label{qed4-gpsi1}
\eea
\end{subequations}
where $L_p= \log \frac{-p^2}{\mu^2}$. Eq.~(\ref{qed4-gpsi1}) shows that the fermion field has an anomalous scaling dimension.
On the other hand, for QED$_3$ we first have to fix $\varepsilon_e=0$ and then take the limit 
$\varepsilon_\gamma \rightarrow 1/2$. The theory is then found to be finite: $Z_A=Z_\psi=Z_\Lambda=1+ \mathcal{O}\left(\al^2\right)$.
The 1-loop expressions for the vacuum polarization and fermion self-energy read:
\begin{subequations}
\label{qed3-1loop}
\bea
&& \Pi_1(q^2) = -\frac{d}{2}\frac{1}{16}\frac{e^2}{\sqrt{-q^2}},
\label{qed3-pi1} \\
&& \Sigma_{V1}(p^2) = - \frac{e^2 a}{16 \sqrt{-p^2}},
\label{qed3-sigma1}
\eea
\end{subequations}
where the coupling constant, $e^2$, has dimension of mass, cf., Eq.~(\ref{dims+params}). All the above results agree with dimensional analysis.

We now focus on RQED$_{4,3}$ ($\varepsilon_e=1/2$ and $\varepsilon_\gamma \rightarrow 0$). 
From Eq.~(\ref{pimunu-1l-1}) and in accordance with dimensional analysis, the vacuum polarization of the layer is finite at one-loop
and the reduced gauge field propagator essentially remains free:
\begin{subequations}
\label{rqed43-gauge}
\bea
&&\Pi_1(q^2) = -\frac{d}{2}\frac{\al}{4 \pi}\frac{\pi^2}{\sqrt{-q^2}},
\label{rqed4.3-pi1} \\
&&Z_A=1+ \mathcal{O}\left(\al^2\right), \qquad \gamma_A = \mathcal{O}\left(\al^2\right)
\label{rqed4.3-ZA} \\
&&\tilde{D}_{\bot}(q^2) = \frac{i}{2\sqrt{-q^2}}\frac{1}{1+\frac{d}{4}\frac{e^2}{16}}.
\label{rqed4.3-D}
\eea
\end{subequations}
The square-root branch cut structure of Eq.~(\ref{rqed4.3-pi1}) is typical of $2+1$-dimensional QFTs, cf., Eq.~(\ref{qed3-1loop}) for QED$_3$.
It yields an imaginary part for $q^2>0$ corresponding to a continuum of single pair (particle-antiparticle) excitations.
Because RQED$_{4,3}$ is renormalizable, the exact $\Pi(q^2)$ also scales as $(-q^2)^{-1/2}$. 
This is in contrast with the case of QED$_3$ where the coupling constant is dimensionful
and higher negative powers of $-q^2$ will appear in the loop expansion, see Eq.~(\ref{qed3-pi2}).
In the case of RQED$_{4,3}$, higher order corrections will therefore affect $\Pi(q^2)$ only through numerical coefficients.
The 2-loop interaction correction numerical coefficient will be derived in the next section, see Eq.~(\ref{rqed4.3-pi2}).

On the other hand, the fermion self-energy of RQED$_{4,3}$ is affected by an ultraviolet pole coming from the first gamma function in the numerator of $G(1,1/2)$, cf., Eq.~(\ref{1loopG}).
It is therefore divergent in accordance with naive power counting. 
Performing the $\varepsilon_\gamma$-expansion in Eq.~(\ref{sigma-1l-1}) yields the following results:
\begin{subequations}
\label{rqed43-fermion}
\bea
&&Z_\psi(\al,a) = 1 - \frac{3a - 1}{3}\,\frac{\al}{4\pi \varepsilon_\gamma} + \mathcal{O}\left(\al^2\right),
\label{rqed4.3-Zpsi} \\
&&\gamma_\psi(\al,a) = \frac{2(3a - 1)}{3}\,\frac{\al}{4\pi} + \mathcal{O}\left(\al^2\right),
\label{rqed4.3-Gpsi} \\
&&-i\Sp S_r(p;\mu) = 1 + \frac{\al}{4\pi}\,\frac{10- 3\tilde{L}_p + 9 a (\tilde{L}_p - 2)}{9} + \mathcal{O}\left(\al^2\right),
\label{rqed4.3-S}
\eea
\end{subequations}
where $\tilde{L}_p = L_p + \log(4)$ contains all the momentum dependence.

Finally, from Eq.~(\ref{vertex-1l-1}) we see that the 1-loop vertex correction is also divergent for RQED$_{4,3}$.
The expression of the corresponding constant reads:
\be
Z_\Lambda = 1 + \frac{3a - 1}{3}\frac{\al}{4\pi \varepsilon_\gamma}\, + \mathcal{O}\left(\al^2\right).
\label{rqed4.3-Zlambda}
\ee
Comparing Eqs.~(\ref{rqed4.3-Zlambda}) and (\ref{rqed4.3-Zpsi}), we see that the Ward identity:
$Z_\Gamma Z_\psi=1$, is satisfied at one-loop (this is also the case for QED$_4$, cf., Eqs.~(\ref{qed4-Zpsi1}) and (\ref{qed4-Zlambda1})
and, trivially, for QED$_3$).
From Eq.~(\ref{Zs-constraint}) we deduce that the charge renormalization constant is determined by the
gauge field renormalization constant.
Together with Eqs.~(\ref{beta_def}) and (\ref{ren_coupling}), this implies that the beta function is
determined by the anomalous dimension of the gauge field:
\be
\beta(\al(\mu)) = -2 \varepsilon_\gamma + \gamma_A(\al(\mu)).
\label{betafunc}
\ee
Because RQED$_{4,3}$ has no anomalous gauge field scaling dimension,
the charge renormalization constant is finite at one-loop:
$Z_\al = 1 + \mathcal{O}\left(\al^2\right)$, and the beta function vanishes:
\be
\beta(\al) = \mathcal{O}\left(\al^2\right).
\ee
%

\section{Perturbation theory: two-loop}
\label{Sec.2loop}

\begin{figure}
    \includegraphics{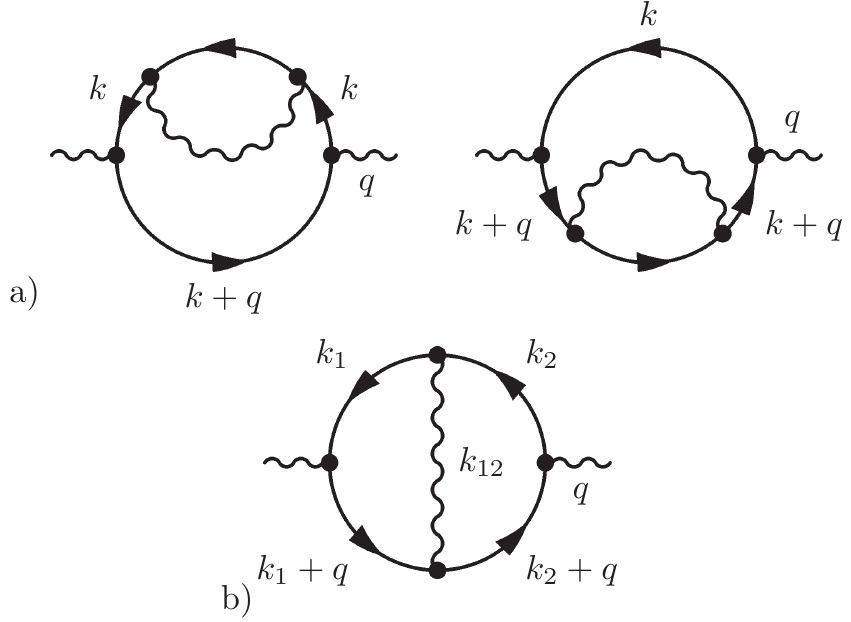}
    \caption{\label{fig:rqed-2loop-polarization}
     2-loop diagrams contributing to the vacuum polarization ($k_{12}=k_1-k_2$).}
\end{figure}

The results we shall derive in this section are expressed in terms of a two-loop massless propagator diagram, cf., Ref.~[\onlinecite{GrozinBook}]:
\bea
&&\int \frac{\D^{d_e}\,k_1}{(2\pi)^{d_e}}\,\frac{\D^{d_e}\,k_2}{(2\pi)^{d_e}}\,\frac{1}{D_1^{\nu_1} D_2^{\nu_2} D_3^{\nu_3} D_4^{\nu_4} D_5^{\nu_5}} =
\label{2loopmassless} \\
&&-\frac{1}{(4\pi)^{d_e}}\,(-q^2)^{d_e-\sum_i\nu_i}\,G(\nu_1,\nu_2,\nu_3,\nu_4,\nu_5),
\nonum \\
&&D_1 = -(k_1+q)^2,~ D_2 = -(k_2+q)^2,~ D_3 = -k_1^2,
\nonum\\
&&D_4 = -k_2^2,~ D_5 = -(k_1-k_2)^2,
\nonum
\eea
where the $\nu_i$s are arbitrary indices labeling the diagram, the power of $-q^2$ follows from dimensionality
and $G(\nu_1,\nu_2, \nu_3,\nu_4,\nu_5)$ is a dimensionless function. An explicit expression for this function
is difficult to derive in the general case.
A class of such massless 2-loop diagrams has been computed by Kotikov [\onlinecite{Kotikov96}].
We will apply these results to the computation of the 2-loop vacuum polarization, cf., Fig.~\ref{fig:rqed-2loop-polarization}.
For simplicity, we will work in the Feynman gauge and set $a=1-\xi=1$ from now on. Moreover, we will assume that
Eq.~(\ref{betafunc}) still holds, i.e. that the Ward identity, $Z_\Gamma Z_\psi=1$, is satisfied at 2 loops.

There are 2 distinct 2-loop diagrams for the gauge-field self energy. The total 2-loop vacuum polarization reads:
\bea
&&\Pi_2^{\mu \nu}(q) = \big(\, g^{\mu \nu} q^2 - q^\mu q^\nu\,\big)\,\Pi_2(q^2),
\nonum \\ 
&&\Pi_2(q^2) = 2\Pi_{2a}(q^2) + \Pi_{2b}(q^2),
\eea
where the first, $a$, diagram appears twice, cf. Fig.~\ref{fig:rqed-2loop-polarization}a.
These contributions are defined as:
\begin{widetext}
\begin{subequations}
\label{2loopdiagrams}
\bea
&&i\Pi_{2a}^{\mu \nu}(q) = 
- \int \frac{\D^{d_e}k}{(2\pi)^{d_e}}\, \Tr \left[ (-ie\gamma^\nu) \frac{i(\Sk +\Sq)}{(k+q)^2}(-ie\gamma^\mu)\frac{i\Sk}{k^2} \left( -i \Sk \Sigma_V(k^2) \right) \frac{i\Sk}{k^2} \right],
\label{pi2loopa}\\
&&i\Pi_{2b}^{\mu \nu}(q) = 
\label{pi2loopb} \\
&&- \int \frac{\D^{d_e}k_1}{(2\pi)^{d_e}}\,\,\frac{\D^{d_e}k_2}{(2\pi)^{d_e}}\, 
\Tr \left[ (-ie\gamma^\nu) \frac{i(\Sk_2 +\Sq)}{(k_2+q)^2}
(-ie\gamma^\al)\frac{i(\Sk_1+\Sq)}{(k_1+q)^2}
(-ie\gamma^\mu)\frac{i\Sk_1}{k_1^2}
(-ie\gamma^\beta)\frac{i\Sk_2}{k_2^2}
\frac{i g_{\al \beta}}{(4\pi)^{\varepsilon_e}}\frac{\Gamma(1-\varepsilon_e)}{(-(k_1-k_2)^2)^{1-\varepsilon_e}}
\right],
\nonum
\eea
\end{subequations}
where $\Sigma_V$ is the one-loop fermion self-energy given by Eq.~(\ref{sigma-1l-1}). 
With the help of the latter, as well as Eq.~(\ref{pi_id}) and the 2-loop trace identity Eq.~(\ref{trace_id-2loop}), the first contribution reads:
\begin{subequations}
\label{2loopdiagrams-a}
\bea
&&\Pi_{2a}(q^2) = -d\, \frac{e^4 \Gamma(1-\varepsilon_e) \left(-q^2\right)^{-2\varepsilon_\gamma-\varepsilon_e}}{(4\pi)^{\frac{d_e+d_\gamma}{2}}}
\frac{4(1-\varepsilon_\gamma -\varepsilon_e)^4}{\varepsilon_\gamma (3- 2\varepsilon_\gamma -2\varepsilon_e)(2-2\varepsilon_\gamma -\varepsilon_e)}\,G_2(\varepsilon_e),
\label{pi2loopa2}\\
&&G_2(\varepsilon_e) = G(1,1-\varepsilon_e)\,G(1,\varepsilon_\gamma).
\label{g2eps}
\eea
\end{subequations}
In Eq.~(\ref{pi2loopa2}), the appearance of the dimensionless function $G_2$, Eq.~(\ref{g2eps}), 
translates the fact that this massless 2-loop diagram is actually a product of 2 massless 1-loop diagrams, cf., Eq.~(\ref{1loopG}).

The second contribution, $b$, is a truly 2-loop diagram, cf., Fig.~\ref{fig:rqed-2loop-polarization}b. After some tedious algebra, it reads:
\begin{subequations}
\label{2loopdiagrams-b}
\bea
\Pi_{2b}(q^2) = &&-d\,\frac{e^4 \Gamma(1-\varepsilon_e) \left(-q^2\right)^{- 2\varepsilon_\gamma - \varepsilon_e}}{(4\pi)^{\frac{d_e+d_\gamma}{2}}}\,
\frac{1-\varepsilon_e - \varepsilon_\gamma}{3-2\varepsilon_e - 2\varepsilon_\gamma}
\left[ G_2(\varepsilon_e)\mathcal{N}_b(\varepsilon_e, \varepsilon_\gamma) + G(1,1,1,1,1-\varepsilon_e) \mathcal{M}_b(\varepsilon_e, \varepsilon_\gamma) \right],
\label{pi2loopb2}\\
\mathcal{N}_b(\varepsilon_e, \varepsilon_\gamma) = &&4 \left[ \varepsilon_e + \varepsilon_\gamma +\frac{2\varepsilon_e}{2-\varepsilon_e-2 \varepsilon_\gamma}
-\frac{2(2-2\varepsilon_e-3\varepsilon_\gamma)}{\varepsilon_\gamma} - \frac{\varepsilon_e (3-\varepsilon_e-3\varepsilon_\gamma)}{(2-\varepsilon_e-2 \varepsilon_\gamma)^2}
\right. \\
&&\left. -\frac{(1-\varepsilon_\gamma)(1-\varepsilon_e - 3\varepsilon_\gamma)(2- 2\varepsilon_e - 3\varepsilon_\gamma)}{\varepsilon_\gamma (1-\varepsilon_e -2\varepsilon_\gamma)
(2-\varepsilon_e -2\varepsilon_\gamma)}
+ \frac{(2+\varepsilon_e + \varepsilon_\gamma)(1-\varepsilon_e - 3\varepsilon_\gamma)(2- 2\varepsilon_e - 3\varepsilon_\gamma)}{\varepsilon_\gamma (1-\varepsilon_e -2\varepsilon_\gamma)} \right],
\nonum \\
\mathcal{M}_b(\varepsilon_e, \varepsilon_\gamma) =
&&-2 \left[ 1 - \frac{\varepsilon_\gamma(1-\varepsilon_\gamma)}{(1-\varepsilon_e -2\varepsilon_\gamma)(2-\varepsilon_e -2\varepsilon_\gamma)}
+ \frac{\varepsilon_\gamma(2+ \varepsilon_e +\varepsilon_\gamma)}{1-\varepsilon_e -2\varepsilon_\gamma} \right].
\eea
\end{subequations}
In Eq.~(\ref{pi2loopb2}), $G_2$ is given by Eq.~(\ref{g2eps}) while the diagram $G(1,1,1,1,1-\varepsilon_e)$ has been computed by Kotikov~[\onlinecite{Kotikov96}] 
and involves a ${}_3 F_2$ hypergeometric function. As quoted by Grozin [\onlinecite{GrozinBook}], this diagram reads:
\bea
G(1,1,1,1,1-\varepsilon_e) &=& 2 \Gamma(1- \varepsilon_e - \varepsilon_\gamma)\Gamma(-\varepsilon_\gamma) \Gamma(\varepsilon_e+2\varepsilon_\gamma) \times 
\label{KotikovG}\\ 
&\times & \left [ \frac{- \Gamma(1- \varepsilon_e - \varepsilon_\gamma)}{(1+\varepsilon_\gamma)\Gamma(2-\varepsilon_e)\Gamma(1-2\varepsilon_e-3\varepsilon_\gamma)}\,
\setlength\arraycolsep{1pt}
\pFq{3}{2}{1, , 2-2\varepsilon_e -2\varepsilon_\gamma, , 1+\varepsilon_\gamma}{2-\varepsilon_e, , 2+\varepsilon_\gamma}{1}
+\frac{\pi \cot \pi (\varepsilon_e+2\varepsilon_\gamma)}{\Gamma(2-2\varepsilon_e -2\varepsilon_\gamma)} \right ].
\nonum
\eea
Eqs.~(\ref{2loopdiagrams-a}) and (\ref{2loopdiagrams-b}) apply to the well known cases of QED$_4$ and QED$_3$.
For the sake of completeness we give the results for QED$_4$ ($\varepsilon_e=0$ and $\varepsilon_\gamma \rightarrow 0$):
\begin{subequations}
\label{qed4-gauge-2loop}
\bea
&&Z_A(\al) = 1 
-\frac{d}{3}\,\frac{\al}{4\pi \varepsilon_\gamma} -
\frac{d \varepsilon_\gamma}{2}\,\left(\frac{\al}{4\pi \varepsilon_\gamma} \right)^2 + \mathcal{O}\left(\al^3\right),
\label{qed4-ZA2}\\
&&D_{\bot}^r(q^2) = \frac{i}{-q^2}
\left[
1 + \frac{d}{3} \left(L_q-\frac{5}{3} \right)\frac{\al}{4\pi}
+ d\left( 4\zeta_3 -\frac{55}{12} + L_q + \frac{d}{9}(L_q-\frac{5}{3})^2 \right) \, \left(\frac{\al}{4\pi} \right)^2\, + \mathcal{O}\left(\al^3\right) \right],
\label{qed4-Dr2}\\
&&\gamma_A(\al) = \frac{2d}{3}\,\frac{\al}{4\pi} + 2d \left(\frac{\al}{4\pi} \right)^2 + \mathcal{O}\left(\al^3\right),
\label{qed4-gammaA}\\
&&\beta(\al) = \frac{2d}{3}\,\frac{\al}{4\pi} + 2d \left(\frac{\al}{4\pi} \right)^2 + \mathcal{O}(\al^3),
\eea
\end{subequations}
where $L_q= \log \frac{-q^2}{\mu^2}$ and the renormalized gauge field propagator has been expanded up to second order in $\al$, see also Ref.~[\onlinecite{GrozinBook}]. 
From Eq.~(\ref{qed4-Dr2}), we notice that in the large-$d$ limit the expansion of the gauge field propagator corresponds to a geometric series.
This is equivalent to a large-$N$ approximation where radiative corrections reduce to the 1-loop vacuum polarization.
On the other hand  QED$_3$ ($\varepsilon_e=0$ and $\varepsilon_\gamma \rightarrow 1/2$) is finite at 2 loops: $Z_A=1+ \mathcal{O}\left(\al^3\right)$.
Adding the 1-loop result, Eq.~(\ref{qed3-pi1}), the total vacuum polarization of QED$_3$ up to 2 loops reads:
\be
\Pi(q^2) = -\frac{d}{2}\frac{1}{16}\frac{e^2}{\sqrt{-q^2}} - \frac{d}{2}\frac{1}{16}\frac{10 -\pi^2}{2\pi^2}\frac{e^4}{-q^2} + \mathcal{O}(e^6),
\label{qed3-pi2}
\ee
where the bare coupling constant, $e^2$, has dimension of mass, cf., Eq.~(\ref{dims+params}).
Focusing now on RQED$_{4,3}$ ($\varepsilon_e=1/2$ and $\varepsilon_\gamma \rightarrow 0$)
we find, in agreement with dimensional analysis as well as the 1-loop results of the last section, that the vacuum polarization of the layer is finite at 2 loops:
\begin{subequations}
\label{rqed43-gauge-2loop}
\bea
&&\Pi_2(q^2) = - \frac{d}{2}\,\left(\frac{\al}{4\pi} \right)^2\,\frac{2\pi^2 (92 - 9\pi^2)}{9 \sqrt{-q^2}},
\label{rqed4.3-pi2} \\
&&Z_A=1+ \mathcal{O}\left(\al^3\right), \qquad \gamma_A = \mathcal{O}\left(\al^3\right)
\label{rqed4.3-ZA2} \\
&&\tilde{D}_{\bot}(q^2) = \frac{i}{2\sqrt{-q^2}} \left[
1 - \frac{d\pi^2}{4}\frac{\al}{4\pi} + \left( -\frac{46d \pi^2}{9} + \frac{d(8+d)\pi^4}{16} \right) \, \left(\frac{\al}{4\pi} \right)^2\ + \mathcal{O}\left(\al^3\right) \right],
\label{rqed4.3-D2}
\eea
\end{subequations}
\end{widetext}
where the reduced gauge field propagator, Eq.~(\ref{rqed4.3-D2}), has been expanded up to second order in $\al$. 
Just as in the case of QED$_4$, we note that in the large-$d$ limit the expansion of the gauge field propagator, 
Eq.~(\ref{rqed4.3-D2}), corresponds to a geometric series. This implies that radiative corrections reduce to the 
1-loop vacuum polarization, in agreement with large-$N$ approximations.
Together with Eq.~(\ref{betafunc}), Eq.~(\ref{rqed4.3-ZA2}) implies the vanishing of the 2-loop beta function of RQED$_{4,3}$:
\be
\beta(\al) = \mathcal{O}(\al^3).
\ee
Adding the 1-loop contribution Eq.~(\ref{rqed4.3-pi1}) to Eq.~(\ref{rqed4.3-pi2}) the total, up to 2 loops, gauge-field self-energy may be written as:
\be
\Pi(q^2) = \Pi_1(q^2) \left( 1 + \al C_\al + \mathcal{O}(\al^2) \right), \quad C_\al = \frac{92-9\pi^2}{18\pi},
\label{Pi2loopCal}
\ee
where $C_\al \approx 0.056$ is an interaction correction numerical constant, cf., discussion below Eq.~(\ref{rqed4.3-pi1}).

\section{Conclusion}
\label{Sec.Conclusion}

A brief account on the computation of radiative corrections in RQED$_{d_\gamma,d_e}$ has been given with a special focus on
electromagnetic current correlations and applications to RQED$_{4,3}$. 
In the latter case, the fermions are strictly $2+1$-dimensional while the gauge field is $3+1$-dimensional. 
Such a theory has been contrasted with usual QED$_4$ and QED$_3$ where both the fermions
and the gauge field live in spaces of the same dimensionality, $3+1$-dimensional and $2+1$-dimensional, respectively.
It can also be contrasted with QED with anisotropic fermion dispersion relation and gauge couplings, 
see Ref.~[\onlinecite{Alexandre2001}], which interpolates between QED$_4$ and RQED$_{4,3}$.
Concerning RQED$_{4,3}$, our main results concern the vanishing of the beta function of the theory up to two loops, 
which implies the scale invariance of the theory up to this order, together with the presence of a non-trivial anomalous
scaling dimension for the fermion field already at one-loop, Eqs.~(\ref{rqed43-fermion}). In the condensed matter literature these results are reminiscent of the
$1+1$-dimensional Tomonaga-Luttinger (TL)  model, see Ref.~[\onlinecite{DiCM-1991+1993}], the beta function of which has been shown to vanish to all orders. 
Actually, RQED$_{4,2}$ is a kind of TL model with additional (four-dimensional) long-range retarded interactions and shares the same feature. 
These $1+1$-dimensional models, together with the Schwinger model, are exact at 1 loop and therefore do not display any non-trivial multi-loop corrections to the polarization diagram 
contrary to their higher dimensional counterparts. As a matter of fact, in the case of RQED$_{4,3}$, the two-loop result allowed us 
to access the interaction correction numerical constant, cf., Eq.~(\ref{Pi2loopCal}) and
Refs.~[\onlinecite{HerbutMishchenko2008}] in the non-relativistic case.
The smallness of this constant is an indication of the fact that, even if the fine structure constant is large, e.g., of the order
of unity, the perturbative approach seems to be justified, at least as far as the 2-loop contribution is concerned.

\acknowledgments

I thank Marc Bellon for helpful discussions.


\end{document}